\documentstyle[12pt,psfig,epsf]{article}

\begin{document}
\title{Extraction of the neutron charge form factor
from the charge form factor of deuteron}

\author{ A.~F.~Krutov\thanks{E-mail:
krutov@ssu.samara.ru}\\
{\small {\em Samara State University, 443011, Samara, Russia
}}\\
V.~E.~Troitsky\thanks{E-mail:
troitsky@theory.sinp.msu.ru}\\
{\small {\em D.V.~Skobeltsyn Institute of Nuclear Physics, 
Moscow State University, 119899, Moscow, Russia }} }
\date{February 19, 2002}
\maketitle

\begin{abstract}
We extract the neutron charge form factor from the charge form
factor of deuteron obtained from
$T_{20}(Q^2)$ data at $0\le Q^2\le$ 1.717 (GeV$^2$).
The extraction is based on the relativistic impulse
approximation in the instant form of the
relativistic Hamiltonian dynamics. Our results (12 new points)
are compatible with existing values of the  neutron charge form
factor of other authors. We propose a fit for the whole set
(35 points) taking into account the data for the slope of the
form factor at $Q^2 = 0$.
\end{abstract}



\noindent

\bigskip

The behavior of the neutron charge form factor
$G^n_E(Q^2)\;,
\;(Q^2 =-q^2\;,\;q$ - the momentum transfer)
is of great importance for the understanding of the
electromagnetic structure of nucleons and nuclei. However,
$G^n_E(Q^2)$ is still known  rather poorly.

As there are no free neutron targets, $G^n_E(Q^2)$ has to be
extracted from the data for composite nuclei, for example 
deuteron or $^3$He \cite{HaD73, MuT81s, JoB91, LuS93, EdM94, Mey94, 
PaA99, OsH99, RoB99, Her99, GoZ01, ToR01, ScS01}. The direct 
measurement of great precision ($\simeq$ 1.5\%) is possible only 
for the slope  $dG^n_E(Q^2)/dQ^2$ at $Q^2=0$, as determined by 
thermal neutron scattering \cite{Kop95}.

While obtaining the information about the neutron from the
scattering data on composite systems one encounters two kinds of
difficulties. First, the results depend crucially on the model
for $NN$ interaction
\cite{ScS01, GaK71, Pla90} .
Second, there exists a dependence on the
relativistic effects, exchange currents, nucleon isobar
states, final state interaction in inelastic channels
etc. \cite{LuS93, ToR01}.
The use of polarized beams and polarized targets in
recent experiments diminishes uncertainties due to those
effects
\cite{JoB91, LuS93, EdM94, PaA99, OsH99, RoB99, GoZ01, ScS01}.

In the present paper the neutron charge form factor is extracted
from the experimental data on the deuteron charge form factor
obtained through polarization experiments on elastic $ed$
scattering \cite{The91, BoA99, AbA00}.
In the JLab experiments \cite{AbA00}
the deuteron charge form factor  is obtained up to
$Q^2$ = 1.717 (GeV$^2$). In this range of momentum transfer
the theoretical description of the polarization tensor
$T_{20}(Q^2)$
depends essentially on the choice of the form of $NN$
interaction and relativistic approach is required.

Our calculations are based on the method of relativistic
Hamiltonian dynamics (RHD) which is widely used in present time.
One can find the description of RHD method in the reviews
\cite{Pol89}
and especially the case of the  deuteron in the reviews
\cite{GaV01}, \cite{GiG01}).
We use our own variant ~\cite{KrT01}, \cite{BaK95}
of the instant form of RHD.  This variant permits to take
correctly into account the relativistic effects in the elastic
{\it ed} scattering in the relativistic impulse approximation
\cite{III}.
The main point of our approach is the construction of the
electromagnetic--current operator for the system of interacting
particles. In our approach this operator is
Lorentz covariant and satisfies the conservation law.

Let us note that, as far as we know, it is for the first time
that the neutron charge form factor is determined from an
analysis of the deuteron charge form factor.

In our approach in the relativistic impulse approximation the
following equation for the deuteron charge form factor takes
place (see
\cite{III} for details):
\begin{equation}
G_C(Q^2) = G_{CC}(Q^2)\left[G^p_{E}(Q^2) + G^n_{E}(Q^2)\right] +
           G_{CM}(Q^2)\left[G^p_{M}(Q^2) + G^n_{M}(Q^2)\right]\;.
\label{Gc}
\end{equation}
Here $G^{p,n}_{E,M}$  are charge and magnetic form factors of
proton and neutron. The fact that nucleons magnetic form factors
enter the Eq. ~(\ref{Gc}) is due to the relativistic effect.

The functions $G_{CC}\;,\;G_{CM}$ in (\ref{Gc})
are given by:
\begin{equation} G_{CC}(Q^2) =
\sum_{l,l'}\int\,d\sqrt{s}\,d\sqrt{s'}\, \varphi^l(s)\,
g^{ll'}_{CC}(s\,,Q^2\,,s')\,\varphi^{l'}(s')\;,
\label{Gcc}
\end{equation}
\begin{equation}
G_{CM}(Q^2) =
\sum_{l,l'}\int\,d\sqrt{s}\,d\sqrt{s'}\, \varphi^l(s)\,
g^{ll'}_{CM}(s\,,Q^2\,,s')\,\varphi^{l'}(s')\;,
\label{Gcm}
\end{equation}
here $\varphi^l(s)$ is the wave function in the sense of RHD
(see \cite{Pol89, KrT01, BaK95}):
\begin{equation}
\varphi^l(s) =
\sqrt[4]{s}\,u_l(k)\,k\;,\quad
k = \frac{1}{2}\sqrt{s - 4\,M^2}\;,\quad
\sum_l\int\,u_l^2(k)\,k^2\,dk = 1\;,
\label{phi}
\end{equation}
$M$ is the nucleon mass, $l=$ 0,2 -- the nucleon angular
momentum in the deuteron,
$u_l(k)$ -- the wave function for the model $NN$  interaction.
The functions $g^{ll'}_{CC}\;,\;g^{ll'}_{CM}$
are given by the following equations
(\ref{gCC})--(\ref{s12}) (note that the same equations
were obtained independently in \cite{AfA98}):
\begin{equation} g^{ll'}_{CC}(s, Q^2, s') = R(s, Q^2, s')\,(s +
s' + Q^2)\,Q^2\, a^{ll'}(s, Q^2, s')\;, \label{gCC}
\end{equation}
\begin{equation}
g^{ll'}_{CM}(s, Q^2, s') =
\frac{1}{M}\,R(s, Q^2, s')\xi(s,Q^2,s')\,
Q^2\,b^{ll'}(s, Q^2, s')\;,
\label{gCM}
\end{equation}
$$
a^{00} = \left(\frac{1}{2}\cos\omega_1\cos\omega_2 +
\frac{1}{6}\sin\omega_1\sin\omega_2\right)\;,\quad
a^{02} = -\frac{1}{6\sqrt{2}}\,
\left(P'_{22} + 2\,P'_{20}\right)\sin\omega_1\sin\omega_2\;,
$$
$$
a^{22} =
\left[\frac{1}{2}\,L_1\,\cos\omega_1\cos\omega_2 +
\frac{1}{24}\,L_2\,\sin(\omega_2 - \omega_1)
+ \frac{1}{12}\,L_3\,\sin\omega_1\sin\omega_2\right]\;,
$$
$$
b^{00} =
\left(\frac{1}{2}\cos\omega_1\sin\omega_2 -
\frac{1}{6}\sin\omega_1\cos\omega_2\right)\;,\quad
b^{02} = \frac{1}{6\sqrt{2}}\,
\left(P'_{22} + 2\,P'_{20}\right)\sin\omega_1\cos\omega_2\;,
$$
$$
b^{22} =
-\,\left[-\frac{1}{2}\,L_1\,\cos\omega_1\sin\omega_2 +
\frac{1}{24}\,L_2\,\cos(\omega_2 - \omega_1) +
\frac{1}{12}\,L_3\,\sin\omega_1\cos\omega_2\right]\;.
$$
$$
R(s, Q^2, s') = \frac{(s+s'+Q^2)}{\sqrt{(s-4M^2) (s'-4M^2)}}\,
\frac{\vartheta(s,Q^2,s')}{{[\lambda(s,-Q^2,s')]}^{3/2}}
\frac{1}{\sqrt{1+Q^2/4M^2}}\;,
$$
$$
\xi(s,Q^2,s')=\sqrt{ss'Q^2-M^2\lambda(s,-Q^2,s')}\;,\quad
\lambda(a,b,c) = a^2 + b^2 + c^2 -2(ab + ac + bc)\;,
$$
$$
L_1 = L_1(s,Q^2,s') = P_{20}\,P'_{20} + \frac{1}{3}P_{21}\,P'_{21} +
\frac{1}{12}P_{22}\,P'_{22}\;,
$$
$$
L_2 = L_2(s,Q^2,s') = P_{21}\left(P'_{22} - 6\,P'_{20}\right) -
P'_{21}\left(P_{22} - 6\,P_{20}\right)\;,
$$
$$
L_3 = L_3(s,Q^2,s') = 2\,P_{21}\,P'_{21} + 4\,P_{20}\,P'_{20} -
P_{20}\,P'_{22} - P_{22}\,P'_{20}\;.
$$
Here
$\omega_1$ and $\omega_2$ are the Wigner spin rotation
parameters:
$$
\omega_1 =
\arctan\frac{\xi(s,Q^2,s')}{M\left[(\sqrt{s}+\sqrt{s'})^2 +
Q^2\right] + \sqrt{ss'}(\sqrt{s} +\sqrt{s'})}\>,
$$
\begin{equation}
\omega_2 = \arctan\frac{
\alpha (s,s') \xi(s,Q^2,s')} {M(s + s' + Q^2)
\alpha (s,s')
+ \sqrt{ss'}(4M^2 + Q^2)}\>,
\label{omega}
\end{equation}
and $\alpha (s,s') = 2M + \sqrt{s} + \sqrt{s'}$.

$P_{2i} = P_{2i}(z)\;,\;P'_{2i} = P_{2i}(z')\;,\; i=0, 1, 2$ --
the Legendre functions:
\begin{equation}
P_{20}(z) =
\frac{1}{2}\left(3\,z^2 -1\right)\;,\quad P_{21}(z) =
3\,z\sqrt{1 - z^2}\;,\quad P_{22}(z) = 3\left(1 - z^2\right)\;.
\label{Leg}
\end{equation}
\begin{equation}
z = z(s,Q^2,s') =
\frac{\sqrt{s}(s' - s - Q^2)}{\sqrt{\lambda(s,-Q^2,s')(s - 4\,M^2)}}\;,\quad
z' = z'(s,Q^2,s') = -\,z(s',Q^2,s)\;.
\label{zz'}
\end{equation}
$\vartheta(s,Q^2,s')=
\theta(s'-s_1)-\theta(s'-s_2)$, $\theta$ is the step function.
\begin{equation}
s_{1,2}=2M^2+\frac{1}{2M^2} (2M^2+Q^2)(s-2M^2)
\mp \frac{1}{2M^2}
\sqrt{Q^2(Q^2+4M^2)s(s-4M^2)}\;.
\label{s12}
\end{equation}
The functions $s_{1,2}(s,Q^2)$ give the kinematically available
region in the plane
$(s,s')$ (see ~\cite{TrS69}).
\begin{equation}
g^{ll'}_{Ci}(s, Q^2, s') = g^{l'l}_{Ci}(s', Q^2, s)\;,\quad i=C,M\;.
\label{iCM}
\end{equation}

Using Eq.(\ref{Gc}) one can write the neutron charge form factor
in the form:
\begin{equation}
G^n_E(Q^2) = \frac{G_C(Q^2)}{G_{CC}(Q^2)} - \frac{G_{CM}(Q^2)}{G_{CC}(Q^2)}
\left[G^p_{M}(Q^2) + G^n_{M}(Q^2)\right] - G^p_{E}(Q^2)\;.
\label{GnE}
\end{equation}

We calculate the nucleon charge form factor in the points
$Q^2$ where the deuteron charge form factor
$G_C(Q^2)$ is measured. In these points the nucleon form factors
$G^p_{E}(Q^2)\;$, $\;G^p_{M}(Q^2)\;$, $\;G^n_{M}(Q^2)$ are 
obtained through the fits of their experimental values. The 
functions $G_{CC}(Q^2)\;,$ $\;G_{CM}(Q^2)$ can be calculated using 
the equations (\ref{Gcc}), (\ref{Gcm}) and some deuteron wave 
functions.

Let us discuss now the problem of choosing the deuteron wave
functions to use for the calculation of
$G_{CC}(Q^2)\;,\;G_{CM}(Q^2)$  (\ref{GnE}).
We investigated the behavior of $T_{20}(Q^2)$
(our determination of $T_{20}(Q^2)$ is the same as in 
\cite{The91})
using different
wave functions\footnote{The details will be published
elsewhere.}: Paris wave functions
\cite{LaL81}, the versions I, II and 93 of the
Nijmegen model \cite{StK94}, charge--dependent version of
Bonn potential \cite{Mac01},
and the wave functions obtained by the relativistic dispersion
variant of the inverse scattering method
\cite{MuT81}. 
As $T_{20}(Q^2)$ for polarized $ed$ scattering depends weakly
on the form of nucleon form factors one can use the experimental
data for $T_{20}(Q^2)$ to choose the most adequate deuteron wave
functions. Fig.1 presents the results of our calculation of
$T_{20}(Q^2)$ with the use of the wave functions
\cite{LaL81, StK94, Mac01, MuT81} and nucleon form factors from
\cite{GaK71} as well as the experimental points from the papers
\cite{The91, BoA99, AbA00, ScB84, Dmi85,
GiH90, FeB96}.

One can see that the best description of
$T_{20}(Q^2)$  is given by the wave functions \cite{MuT81}.

Our estimations show that other wave functions
(e.g. used in
\cite{LeP00, AlK01}) also give poorer description of
$T_{20}(Q^2)$ than the wave functions
\cite{MuT81}.

Let us emphasize that the wave functions
\cite{MuT81}
were obtained more than 20 years ago and so no possible fitting
reasons for $T_{20}(Q^2)$ could influence the choose. These
wave functions used in the relativistic calculation of the
function
$A(Q^2)$ give the correct behavior up to
$Q^2\;\simeq$ 3 (GeV$^2$).

It seems to us that the validity of the wave functions 
\cite{MuT81} is due to the fact that they are "almost model 
independent":  no form of $NN$ interaction Hamiltonian is used.  
The
wave functions \cite{MuT81} were obtained in the frame of the
potentialless approach to the inverse scattering problem
(see for the details \cite{Tro94}). They are given by the
dispersion type integral directly in terms of the experimental
scattering phases and the mixing parameter for
$NN$ scattering in the $^3S_1-^3D_1$ channel.

In the Eq.(\ref{GnE}) we use for the nucleon form factors
$G^p_{E}(Q^2)\;,\;G^p_{M}(Q^2)\;,\;G^n_{M}(Q^2)$ one
(with the best $\chi^2$) of the fits of the recent paper
\cite{Lom01} --  DRN--GK(3).

The results of our calculations of the neutron charge form
factor in the points where the deuteron charge form factor is
measured are given in the Table 1 (see also Fig.2).

The accuracy of our calculations are determined by the accuracy of
measurements of charge deuteron form factor \cite{The91, BoA99, AbA00} and
nucleon form factors which are the folowing at $Q^2\;\le$ 1.717
(GeV$^2$): for $G^p_{E}(Q^2)$ 1--10\% \cite{HaD73, MuS74, WaF94, AnB94}, 
for $G^p_{M}(Q^2)$ 1--3\% \cite{HaD73, WaF94, AnB94, BaB73, BoK90}, 
for $G^n_{M}(Q^2)$ 1--10\%. \cite{LuS93, MaF93, GaA94, AnB98, XuD00}.

We obtain the first three points at low momentum transfer from
the data for the deuteron charge form factor given in the
paper \cite{BoA99}. In this range of momentum transfer the behavior
of the deuteron charge form factor
and so $G_E^n(Q^2)$
do not depend on the choose of the wave functions 
\cite{LaL81, StK94, Mac01, MuT81}.

The first point at $Q^2\simeq$ 0.16 (GeV$^2$) is almost the same
as in \cite{JoB91}, however, our errors are much smaller.
The second and the third points are compatible (within the
experimental errors) with the
points of \cite{PaA99, OsH99, GoZ01}.
Our point \# 7 is in fact the same as in
\cite{HaD73} but our error is larger.

Our values of $G_E^n$ in other points (at $Q^2\ge$ 1 (GeV$^2$)
are strictly positive. This result differs from e.g. the results
of the paper \cite{LuS93}
consistent with $G_E^n=0$. Let us note that our errors
at $Q^2\;\ge\;$ 1 (GeV$^2$) are sufficiently small, smaller
than, e.g. in
\cite{HaD73,LuS93}.

Our values \# 4--8 are extracted from the values of charge deuteron form
factor of the two different works \cite{The91, AbA00}. The results of these
works are in rather poor agreement with each other in the region
of the first dip. So the values of \# 4--8 of $G_E^n$ are not
well  determined in the present work. One needs
additional experiments in this region.

It is now interesting to fit all the existing values of neutron
charge form factor
(\cite{HaD73, JoB91, LuS93, EdM94, Mey94, PaA99, OsH99, RoB99,
Her99, GoZ01} and Table 1).
We use for the fitting the following function
(see \cite{GaK71} and the review \cite{GaV01})
with two parameters
$a$ and $b$:
\begin{equation} G^n_E(Q^2) =
-\,\mu_n\,\frac{a\,\tau}{1 + b\,\tau}\,G_D(Q^2)\;,\quad G_D(Q^2)
= \left(1 + \frac{Q^2}{0.71}\right)^{-2}\;,\quad \tau =
\frac{Q^2}{4\,M^2}\;.
\label{fit}
\end{equation}
The neutron magnetic moment $\mu_n$ = -1.91304270(5)
\cite{Gro00}. $Q^2$ in $G_D(Q^2)$
is given in (GeV$^2$).

We obtain the parameter $a$
from the slope of the neutron charge form factor at $Q^2 = 0$
\cite{Kop95, Lom01}:
\begin{equation}
\left.
\frac{dG^n_E}{dQ^2}\right|_{Q^2=0} = 0.0199\pm 
0.0003\;\hbox{fm}^2\;.  
\label{slope} 
\end{equation} 
The fitting 
of the slope (\ref{slope}) gives $a$=0.942 with the accuracy 
$\approx$  1.5\%.

This value of $a$ gives the slope of $G_E^n(Q^2)$ at $Q^2 = 0$
which is measured directly in the experiment.

The parameter $b$ is fitted using the $\chi^2$ criterion.
If we use all the 35 points we obtain
$b$ = 4.61 with $\chi^2$ = 69.0.
Note that the fit DRN--GK(3) \cite{Lom01}
of 23 points has $\chi^2$ = 63.9.

If we exclude the points \# 4--8 then the 30--point fitting
gives $b=$ 4.62 with $\chi^2$ = 61.5. As the errors of these
points are large this fitting differs from the previous one
slightly.

Let us note that our fitting for 23 points of the papers
\cite{HaD73, JoB91, LuS93, EdM94,
Mey94, PaA99, OsH99, RoB99, Her99, GoZ01}
(not taking into account our points) gives
$b$ = 4.69 with $\chi^2=$ 57.7. The two curves lie near one
another (see Fig.2) so our points are consistent with the known
points of other authors.

The results of fitting, the experimental points
\cite{HaD73, JoB91, LuS93, EdM94, Mey94, PaA99, OsH99, RoB99, Her99, GoZ01},
as well as our new points are shown on the Fig.2. The points
\# 5 and \# 6 are out of the figure.

To summarize,

1) We extract new points for the neutron charge form factor from
the experimental data for the deuteron charge form factor. The
obtained values are consistent with the known values of other
authors.

2) We perform the fitting for 35 values of the neutron charge
form factor including our points. The fit has the form
(\ref{fit}) with $a$ = 0.942, $b$ = 4.61.

This work was supported in part by the Program "Russian
Universities -- Basic Researches" (grant \# 02.01.28).

\pagebreak

Table 1. The values of $G^n_E(Q^2)/G_D(Q^2)$ obtained in the
present paper. The values of the deuteron charge form factor
used for the extraction of $G^n_E(Q^2)$ are also given.

\begin{center}
\begin{tabular}{|c|c|c|c|c|}
\hline
\# of points & $Q^2$ (GeV$^2$) & $G_C(Q^2)$      & Ref.&$G^n_E(Q^2)/G_D(Q^2)$\\
\hline 
1 & 0.160 & 0.163$\pm$0.017&\cite{BoA99}&        0.076$\pm$0.116   \\
\hline 
2 & 0.215 &0.100$\pm$0.012 &\cite{BoA99}&        0.052$\pm$0.129    \\
\hline
3 & 0.303 & 0.035$\pm$0.020 &\cite{BoA99}&        -0.234$\pm$0.401    \\
\hline
4 & 0.556 & (0.127$^{+0.047}_{-0.056})\cdot$10$^{-1}$&\cite{The91} &  1.23$\pm$0.92\\ 
\hline 
5 & 0.651 &(-0.117$\pm$0.162)$\cdot$10$^{-2}$           &\cite{AbA00} & -2.61$\pm$1.65\\ 
\hline 
6 & 0.693 &(0.166$^{+0.161}_{-0.142})\cdot$10$^{-2}$   &\cite{The91} &-4.52$\pm$5.10\\ 
\hline 
7 & 0.775 &(-0.253$\pm$0.063)$\cdot$10$^{-2}$           &\cite{AbA00} &0.677$\pm$0.361\\ 
\hline 
8 & 0.831 &(-0.147$^{+0.106}_{-0.104})\cdot$10$^{-2}$  &\cite{The91} &-0.140$\pm$0.432\\ 
\hline 
9 & 1.009 &(-0.396$\pm$0.028)$\cdot$10$^{-2}$           &\cite{AbA00} &0.389$\pm$0.107\\ 
\hline 
10& 1.165 &(-0.348$\pm$0.031)$\cdot$10$^{-2}$           &\cite{AbA00} &0.259$\pm$0.131\\ 
\hline 
11& 1.473 &(-0.310$^{+0.053}_{-0.061})\cdot$10$^{-2}$  &\cite{AbA00} &0.405$\pm$0.263\\ 
\hline 
12& 1.717 &(-0.194$^{+0.036}_{-0.052})\cdot$10$^{-2}$  &\cite{AbA00} &0.174$\pm$0.294\\ 
\hline 
\end{tabular}
\end{center}

\pagebreak

Figures capture.\\[1cm]

Fig.1. Data and the results of calculation of
the deuteron polarization tensor
$T_{20}(Q^2)$ for the elastic
$ed$ -- scattering with the use of the nucleon form factors
from the paper \cite{GaK71} and different wave functions.
The experimental points are:
open circles -- \cite{ScB84}, open squares --
\cite{FeB96}, open  triangles -- \cite{GiH90},
filled circles -- \cite{The91},
filled squares -- \cite{BoA99}, filled  diamonds
-- \cite{AbA00}, filled  triangles   -- \cite{Dmi85}.
Curves: solid -- Nijmegen--II \cite{StK94},
dashed  -- \cite{MuT81}, dotted  -- \cite{LaL81},
dash--dotted -- Nijmegen--I \cite{StK94},
dash--dotted--dotted -- \cite{Mac01}.\\[1cm]

Fig.2. The experimental values and the results of fitting for
the neutron charge form factor.
The experimental points:
bold cross -- \cite{EdM94},
open  bold  diamonds -- \cite{GoZ01},
open  up triangles -- \cite{OsH99},
open circles  -- \cite{LuS93},
open down triangles   -- \cite{Mey94},
open   stars -- \cite{Her99},
filled circles -- \cite{PaA99},
filled  diamonds  -- \cite{RoB99},
filled up triangles -- \cite{JoB91},
filled  stars -- \cite{HaD73},
filled squares  -- the present work.
The points \# 5 and \# 6 are out
of the figure.
The curves :  solid --
the result of fitting of 35 experimental points (including
our points of the Table 1) using the equation
(\ref{fit}) ($a$ = 0.942, $b$
= 4.61 with $\chi^2$ = 69.0),  dashed  --
the result of fitting of 23 points of other authors
($a$ = 0.942, $b$ = 4.69 with $\chi^2$ = 57.7).

\pagebreak
{~}

\vspace{4cm}

\begin{figure}[htbp]\vspace*{-3.0cm}
\epsfxsize=0.9\textwidth
\centerline{\psfig{figure=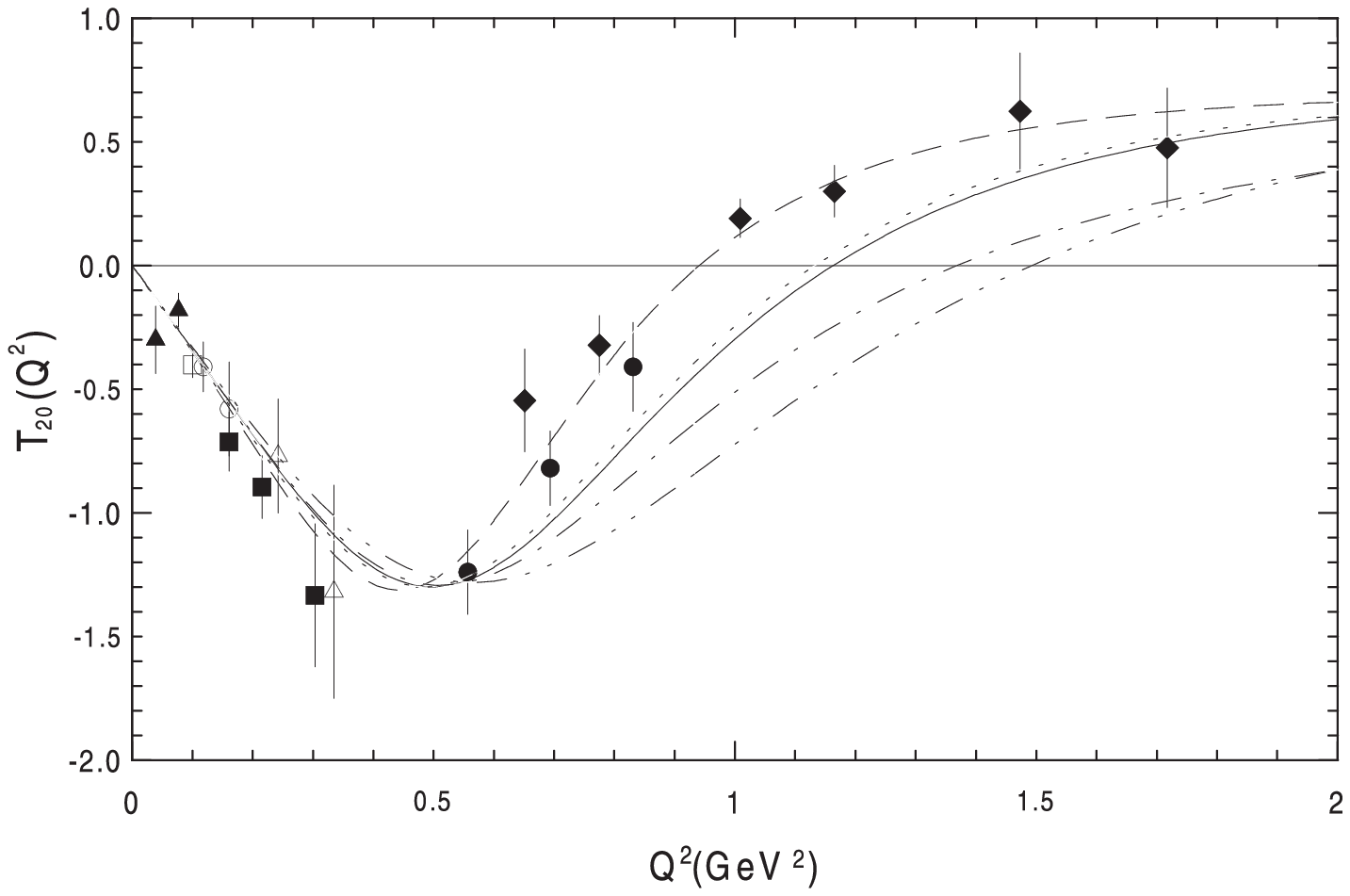,height=10cm,width=18cm}}
\vspace{2cm}
\begin{center}
{FIG.1.}
\end{center}
\vspace*{-2.5cm}
\label{FIG.1.}
\end{figure}

\pagebreak
{~}

\vspace{4cm}

\begin{figure}[htbp]\vspace*{-3.0cm}
\epsfxsize=0.9\textwidth
\centerline{\psfig{figure=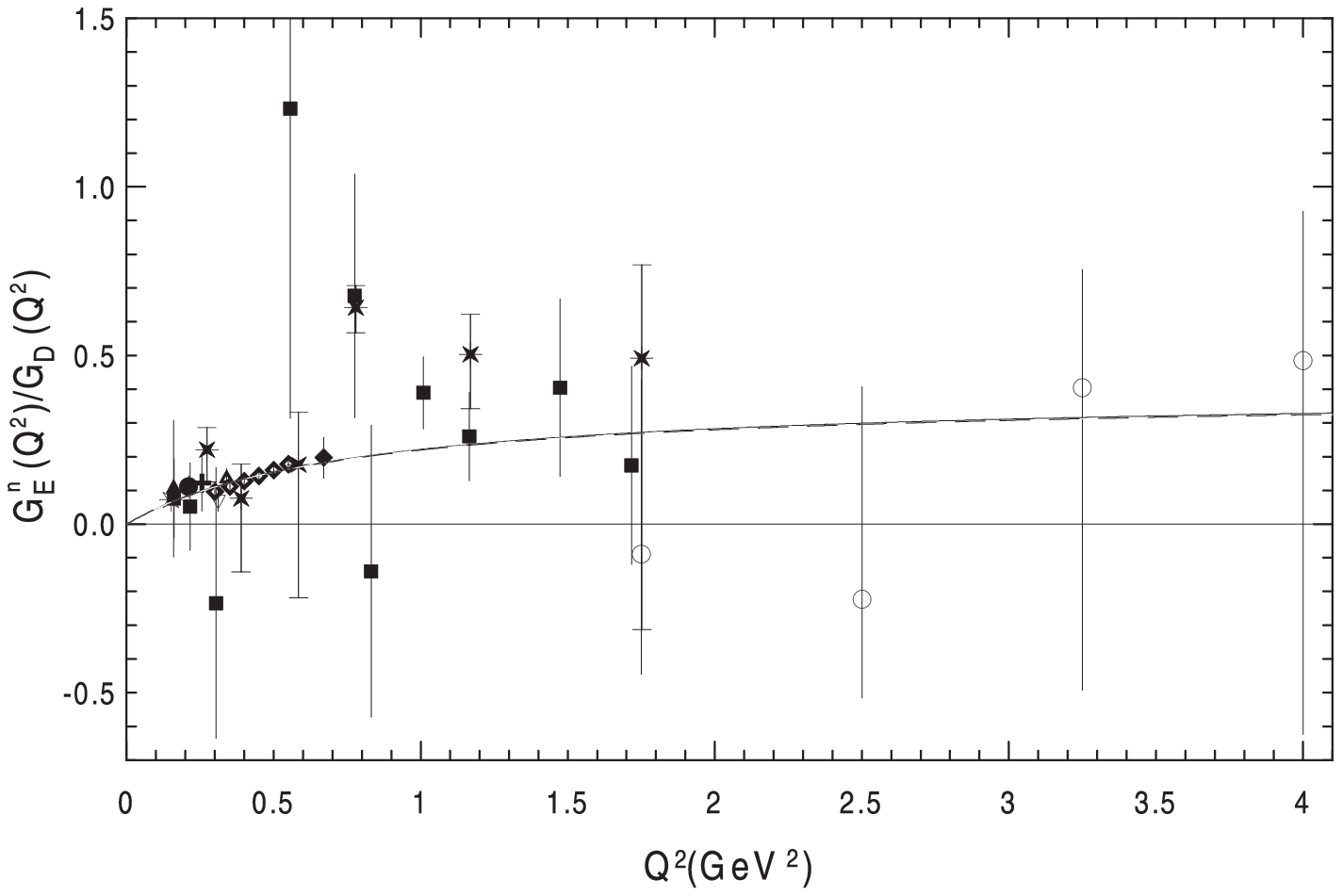,height=10cm,width=18cm}}
\vspace{2cm}
\begin{center}
{FIG.2.}
\end{center}
\vspace*{-2.5cm}
\label{FIG.2.}
\end{figure}

\end{document}